\documentclass[prl,twocolumn,showpacs,amsmath,amssymb]{revtex4-1}
\usepackage{graphicx}       
\usepackage{mathtools}
\usepackage{epsfig}
\usepackage{xcolor}
\usepackage{dcolumn}        
\usepackage{ulem}
\newcommand{\qvec}{{\bf q}}
\newcommand{\beq}{\begin{equation}}
\newcommand{\eeq}{\end{equation}}
\newcommand{\beqa}{\begin{eqnarray}}
\newcommand{\eeqa}{\end{eqnarray}}

\begin{document}
\title{Disorder-driven dissipative quantum criticality as a source of strange metal behavior}
\author{M. Grilli$^{1,2}$, C. Di Castro$^1$, G. Seibold$^3$, S. Caprara$^{1,2}$}
\affiliation{$^1$Dipartimento di Fisica, Universit\`a di 
Roma ``La Sapienza'', P.$^{le}$ Aldo Moro 5, 00185 Roma, Italy}

\affiliation{$^2$ISC-CNR, Unit\`a di Roma ``Sapienza''}

\affiliation{$^3$ Institut f{\"u}r Physik, BTU Cottbus-Senftenberg - PBox 101344, D-03013 Cottbus, Germany} 

\begin{abstract}
{The strange metal behavior, usually characterized by a linear-in-temperature ($T$) resistivity, is a still unsolved mystery in solid-state
physics. Usually it is associated with the proximity to a quantum critical point (a second order transition at temperature $T=0$) focusing 
on the related divergent order parameter correlation length. Here, we propose a paradigmatic shift, focusing on a divergent characteristic 
time scale due to a divergent dissipation acting on the fluctuating critical modes, while their correlation length stays finite. To achieve 
a divergent dissipation, we propose a mechanism based on the coupling between a local order parameter fluctuation and 
electronic diffusive modes, that accounts both for the linear-in-$T$ resistivity and for the logarithmic specific heat versus 
temperature ratio $C_V/T\sim \log(1/T)$, down to low temperatures. 
}
\end{abstract}
\date{\today}
\maketitle

{\it --- Introduction ---}
Although the metallic state is usually well described by Landau's Fermi Liquid (FL) theory, there are many systems in which 
the metallic properties are anomalous, with extended regions of their phase diagram displaying a strange metal behavior \cite{varma-2002,stewart,mackenzie,paschen-2022}. 
The most well-known examples occur in heavy fermion systems in the proximity of quantum critical points (QCPs), i.e., near 
zero-temperature second-order phase transitions, where the uniform metallic state is unstable towards some ordered state 
for some critical value $x_{\mathrm c}$ of a tuning parameter $x$, or in high-temperature superconducting 
cuprates above the optimal superconducting critical temperature (\cite{legros-2019, greene-2020} and references therein). More recent examples are found in iron-based 
superconductors \cite{Walmsley} and twisted bilayer graphene \cite{cao-2020}. The most prominent feature of the 
strange metal behavior is a linear-in-$T$ resistivity without any saturation up to the highest temperatures. Most interestingly, 
this behavior often starts in the vicinity of a QCP with short-ranged two-dimensional (2D) order parameter fluctuations (OPFs) 
and a logarithmic $C_V/T$ ratio ($C_V$ being the specific heat) \cite{stockert-1998,michon-2019}, while other power-laws occur when 
fluctuations are three-dimensional \cite{stewart}. In this work we will focus on 2D systems like cuprates or pnictides, and 
discuss the role of dimensionality in the concluding remarks.

Although some theories for the violation of the FL behavior do not rely on an underlying criticality 
\cite{anderson,kastrinakis,faulkner-2010,hartnoll-2015,sachdev-syk}, the most common interpretations of the strange metal behavior 
rest on the idea that abundant OPFs in the quantum critical region $x\approx x_{\mathrm c}$ may be sufficient to mediate strong 
effective interactions that spoil the Landau quasiparticle stability and create the non-FL state. This scenario 
can be realized in different ways, depending on the nature of the ordered phase, which can be antiferromagnetic \cite{abanov,rosch}, 
charge density wave \cite{cdg-1995}, nematic \cite{metzner-2003}, loop-current \cite{varma-2020}, or can 
have a local character \cite{qimiaosi-2003,coleman-2000,dumitrescu-2021}.

In a previous work \cite{seibold-2020}, we showed that in cuprates charge density fluctuations (CDFs) have a low enough energy $E$ to 
be semiclassical in character (i.e., the Bose function ruling their statistics can be approximated by $T/E$) and are local enough 
(i.e., they involve a sufficiently broad range of momenta) to account for nearly isotropic scattering 
as phenomenologically required by the Marginal Fermi Liquid theory \cite{varma-1989} and recently observed \cite{grissonnanche-2021}. These two ingredients 
are enough to account for the linear-in-$T$ resistivity observed slightly above optimal doping. In subsequent work 
\cite{caprara-2022,mirarchi} we found that a large dissipation of the OPFs may extend the regime of their semiclassical behavior thereby
accounting for a linear-in-$T$ resistivity down to the lowest temperatures and for the observed logarithmic
divergence of the $C_V/T$ ratio. The question remained open about the microscopic mechanisms inducing the required increase of 
dissipation. The present work addresses precisely this issue and provides a possible explanation in terms of coupling between the OPFs 
and the diffusion modes of electrons.

\begin{figure}
  \includegraphics[angle=0,scale=0.18]{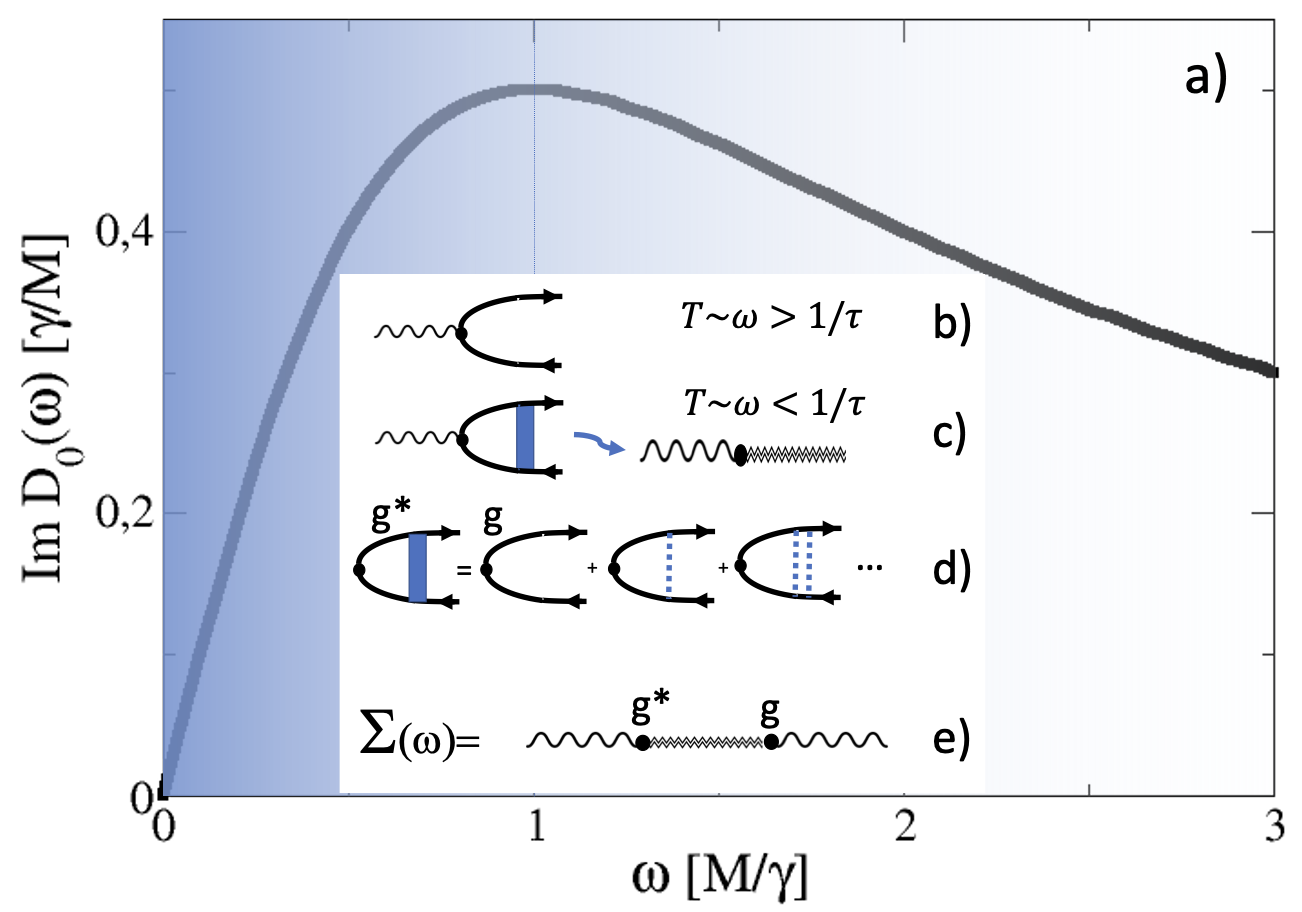}
  \caption{Spectral density of the OPF propagator (a) and sketch of
    the coupling (b-d) between the OPFs (wavy lines) and the p-h diffusive modes (zigzag lines). The solid 
dots represent their effective coupling $g$ entering in Eq.\,(\ref{action}). (b) High-energy regime in which the OPF 
decays into a ballistic p-h pair; (c) low-energy regime in which the OPF decays into a diffusive p-h pair. The blue rectangle represents the 
infinite series of elastic scattering on quenched impurities represented by dotted lines in (d); (e) self-energy diagram for the exact 
solution of the model.
}
\label{fig-ph-decay}
\end{figure}

{\it {--- The scenario ---}}
We consider a regime in which the OPFs have a rather short correlation length, $\xi/\lambda \sim 1-2$ ($\lambda$ is their
characteristic wavelength), by requiring that the system is at a finite distance from the QCP at $x=x_{\mathrm c}$. In this regime the 
fluctuations are largely independent from each other and have a nearly local character, so that they can be represented by a local 
field at the origin, $\Phi({\bf R}=0)$. Accordingly the propagator of these fluctuations has the typical form of an overdamped oscillator 
\begin{equation}
{\cal D}_0(\omega_n)
=\left(M+\gamma |\omega_n|\right)^{-1},
\end{equation}
where $\omega_n$ is the Matsubara frequency, $\gamma$ is a dimensionless parameter measuring the damping strength
due to the decay of the fluctuations into particle-hole (p-h) pairs (Landau damping), the energy 
scale $M=\nu \xi^{-2}$ stays finite, and $\nu$ is an electron energy scale
(we adopt units such that the Planck constant $\hbar$ and the Boltzmann constant $k_{\mathrm B}$ are set equal to 1, so that 
angular frequencies, energies, and temperatures have the same units).
By analytically continuing to real frequencies, $\mathrm i\omega_n \to \omega+\mathrm i0^+$, one can obtain the  spectral density 
of the OPFs, which is broad and peaked at $\omega \approx M/\gamma$, as depicted in Fig.\,\ref{fig-ph-decay}(a).

Depending on the typical energy $M/\gamma$ of the decaying fluctuation, the particle and the hole can propagate as ballistic 
particles when their energy (of order $M/\gamma$) is larger than $1/\tau$, the elastic scattering rate of the charge carriers on quenched 
impurities [Fig.\,\ref{fig-ph-decay}\,(b)]. On the other hand, when the fluctuation has a lower energy, $M/\gamma<1/\tau$, a new 
decay channel opens, with electrons having a diffusive character if $T<1/\tau$ (while for $T>1/\tau$ the electrons and the holes are 
in a quasi-ballistic regime, but the OPFs remain in a classical regime). In this latter case, one can think the nearly local OPF
to decay into a p-h diffusive mode [Fig.\,\ref{fig-ph-decay}\,(c)].
We will show that this diffusive decay channel of the OPFs triggers the growth of $\gamma$ and creates 
the condition to extend the strange metal behavior down to the lowest temperatures and a low-temperature logarithmic growth of the specific 
heat ratio $C_V/T$. Our work therefore provides a concrete microscopic ground for the phenomenological strange metal scenario 
described in Ref.\,\onlinecite{caprara-2022}.

In the standard theory of disordered electron systems \cite{localization,CDKL-1986}, a diffusive collective mode
is obtained by a ladder resummation of impurity scattering events [the dotted lines in 
Fig.\,\ref{fig-ph-decay}\,(d)], so that the  density-density response function takes the form of a diffusive pole
\beq
{\chi}({\bf q},\omega_n)=\langle\rho({\bf q},\omega_n)\rho(-{\bf q},\omega_n)\rangle=\frac{N_0Dq^2}{Dq^2+|\omega_n|}, 
\eeq
where $\bf q$ is the wave vector, $q\equiv|\bf q|$, $D$ is the diffusion constant, and $N_0$ the quasiparticle density of states 
at the Fermi level. These density fluctuations keep their singular diffusive form as long as their typical energy scales
are larger than $T$ and smaller than the elastic scattering rate on quenched impurities $1/\tau$.

To describe an equilibrium situation, where an OPF decays into diffusing p-h pairs, which in turn form back an 
OPF, we introduce a phenomenological coupling $g$ between an OPF (centered at ${\bf R}=0$) and the diffusive density fluctuation
\begin{equation}
{\cal S}_{coupl}=gT\sum_n\Phi({\bf R}=0,\omega_n)\sum_{\bf q} \rho(\qvec,\omega_n). \label{action}
\end{equation}
This coupling describes the possible decay of an OPF into a diffusive mode and vice versa giving rise to an 
exactly solvable model. The coupling between OPFs and diffusive modes dresses the OPF propagator with the self-energy graphically 
represented in Fig.\,\ref{fig-ph-decay}\,(e),
\begin{eqnarray}
  \Sigma(\omega_n)&=&g^2N_0 \int_{Q_{\mathrm{min}}}^{Q_{\mathrm{max}}} \frac{\mathrm d^2 q}{4\pi^2}   
  \frac{Dq^2}{Dq^2+|\omega_n|} \nonumber  \\
  &=& \frac{g^2N_0}{4\pi D} \int_{\mathrm{min}\,(T,\Lambda_{\mathrm{max}})}^{\Lambda_{\mathrm{max}}} \mathrm d(Dq^2) 
  \left(1-\frac{|\omega_n|}{Dq^2+|\omega_n|} \right) \nonumber \\
  &=& \delta M-|\omega_n|\delta\gamma. \label{self-en}
\end{eqnarray}
As usual, the upper momentum cutoff in the diffusion processes is given by the inverse mean free path $Q_{\mathrm{max}}=\ell ^{-1}$, 
which can then be translated into an energy cutoff for the diffusive modes $\Lambda_{\mathrm{max}}\equiv D Q_{\mathrm{max}}^2=1/\tau$. 
The lower cutoff is instead provided by the temperature $T$, as long as $T<1/\tau$. 
The first term in Eq.\,(\ref{self-en}) is a finite correction to the energy scale $M$, which is immaterial in the forthcoming discussion.
Expanding to first order in $|\omega_n|$ the last term in Eq.\,(\ref{self-en}), one obtains a correction to the dissipation 
coefficient $\gamma$,
\beq
\delta\gamma=\gamma-\gamma_0=A \log \mathrm{max}\,[(\tau T)^{-1},1],
\label{log-gamma}
\eeq
where $\gamma_0$ is the damping coefficient in the absence of coupling to diffusive modes and $A\equiv g^2N_0/(4\pi D)$. This result 
remarkably shows that the diffusive channel induces a logarithmic increase of the dissipation parameter $\gamma$ when $T$ decreases.
Since within a phenomenological approach it was previously shown that this leads to the same logarithmic divergence of $C_V/T$ 
\cite{caprara-2022,mirarchi}, this 
result provides a microscopic mechanism accounting for this divergence {\it without} any divergence of the correlation length $\xi$.
This naturally raises the issue of the role of the nearby QCP. In particular, one can notice that Eq.\,(\ref{log-gamma}) does not 
explicitly involve the parameter $x$ tuning the proximity to the QCP, nor the correlation length characterizing the OPFs. We therefore 
need to equip our microscopic model with phenomenological assumptions to determine the range in $x$ where the above diffusive decay 
channel becomes effective. First of all, we consider the condition that, when the OPF has a characteristic energy $M/\gamma_0>1/\tau$, 
it can only decay in ballistic p-h pairs and therefore $g=0$. Since the short-range fluctuations are the 
2D precursors  of the nearby  QCP, the correlation length will increase for $x$ approaching $x_{\mathrm c}$ and the decay in 
diffusive p-h pairs sets in when the tuning parameter of criticality $x$ is such that 
$\nu\xi^{-2}\approx M_0(x-x_{\mathrm c})<\gamma_0/\tau$, i.e., $x<x_{\mathrm{DMD}}\equiv x_{\mathrm c}+\gamma_0/(\tau M_0)$
(DMD stands for diffusive mode decoupling). This sets the maximum distance from the QCP above which $\gamma\approx \gamma_0$.
On the other hand, our arguments (nearly independent OPFs, short correlation length $\xi$) fail when one approaches
the QCP, where the physics is ruled by a diverging correlation length $\xi$ and the standard Hertz-Millis 
picture \cite{hertz-1976,millis-1993} is recovered. 
Therefore, we assume that the diffusive modes decouple from the OPFs for $(x_{\mathrm c}<)\,x<x_{\mathrm{QCR}}$ (QCR stands for quantum 
critical regime), where $g=0$. Then, Eq.\,(\ref{action}) only holds in range $x_{\mathrm{QCR}}<x<x_{\mathrm{DMD}}$.

\begin{figure}
\includegraphics[angle=0,scale=0.3]{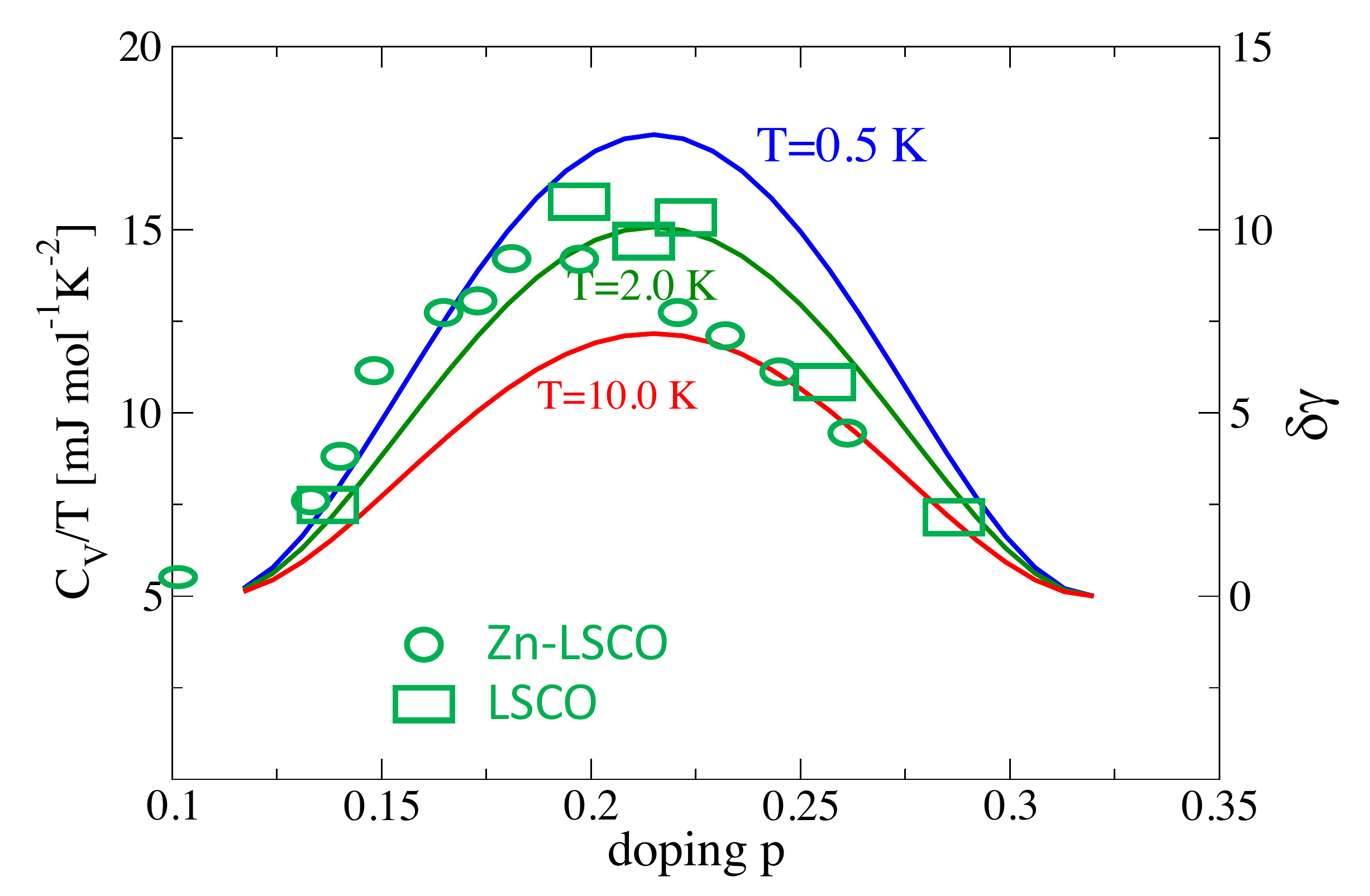}
\caption{(left y axis) The green rectangles or circles
report the electron specific heat experimentally measured in  La$_{2-x}$Sr$_x$CuO$_4$ \cite{girod-2021} and Zn-doped 
La$_{2-x}$Sr$_x$CuO$_4$ \cite{momono-1994}; (right y axis) Dissipative parameter correction $\delta \gamma$, Eq.\,(\ref{log-gamma}), for
$\alpha=18.6$, $\frac{1}{\tau}=500$\,K, $p_{\mathrm{QCR}}=0.11$, and $p_{\mathrm{DMD}}=0.32$. These values have been chosen for
a closer comparison of the $\delta \gamma$ shape with the specific heat anomaly.}
\label{fig-gamma-p}
\end{figure}

{\it {--- The cuprates: specific heat ---}}
For the sake of concreteness, we now specialize our discussion to the case of cuprates, where a charge density wave QCP occurs
near optimal doping, at a critical doping $p_{\mathrm c}$ hidden under the superconducting dome \cite{cdg-1995,andergassen-2001,badoux-2016}. 
We therefore identify the parameter $x$ with doping $p$ and the OPFs with short-ranged CDFs,
recently discovered in resonant X-ray spectroscopy \cite{arpaia-2019}. To implement the constraint that $g\ne 0$ only for 
$p_{\mathrm{QCR}}<p<p_{\mathrm{DMD}}$, we phenomenologically
impose in Eq.\,(\ref{log-gamma}) a doping dependence
\begin{eqnarray}
A(p) \approx \alpha\left[\frac{(p-p_{\mathrm{QCR}}) (p_{\mathrm{DMD}}-p)} {p_{\mathrm{QCR}}\,p_{\mathrm{DMD}}}\right]^2,  \label{avsp}
\end{eqnarray}
for $p_{\mathrm{QCR}}<p<p_{\mathrm{DMD}}$, and $A(p)=0$, otherwise, where $\alpha$ is a suitable parameter.
In the above interval of $p$,  for $T<T_{\mathrm{DMD}}\equiv M/\gamma_0\equiv\nu\xi^{-2}/\gamma_0 \le 1/\tau$ 
the additional diffusive channel is open and $\gamma$ increases, thus lowering $M/\gamma$. 
Fig.\,\ref{fig-gamma-p} describes the behavior of $\gamma(p)$ for various temperatures $T=10$\,K, $2$\,K, and $0.5$\,K. While the 
shape arises from the choice of the doping dependence of $A(p)$ in Eq.\,(\ref{avsp}), the temperature 
dependence follows Eq.\,(\ref{log-gamma}), reproducing the logarithmic behavior 
observed for $C_V/T$ in Refs.\,\onlinecite{michon-2019,girod-2021} and phenomenologically
discussed in Refs.\,\onlinecite{caprara-2022,mirarchi}. Notice that this logarithmic temperature dependence is therefore due to $\delta \gamma$
rather than the logarithmic dependence of the specific heat from the correlation length. Reasonable values of the limiting
control parameters, of the coefficient $A$, of the disorder $1/\tau$ and the temperatures are simply chosen for an easier comparison
between $\delta \gamma$ and $C_V/T$ to which it is proportional \cite{caprara-2022}.
It is worth emphasizing that the $C_V/T$ variation is not unique among 
the cuprates: while it seems to diverge at a specific doping (or in a quite narrow doping range) in Eu-LSCO and Nd-LSCO \cite{michon-2019}, 
it displays instead a broad maximum in LSCO and Bi-2201 \cite{girod-2021,momono-1994}. This variety of behavior is not limited to
specific heat data but corresponds to the possible occurrence of strange metal behavior either in narrow or broad intervals of 
the tuning parameter \cite{stewart,pfeiderer-2007,hussey-2021,hussey-2022}.
Within the present scenario the doping interval  $p_{\mathrm{QCR}}<p<p_{\mathrm{DMD}}$ is non universal and it can be comparatively broader 
or narrower in different cuprates, while the relevant non-trivial result is the logarithmic temperature dependence of $C_V/T$.

\begin{figure}
\includegraphics[angle=0,scale=0.25]{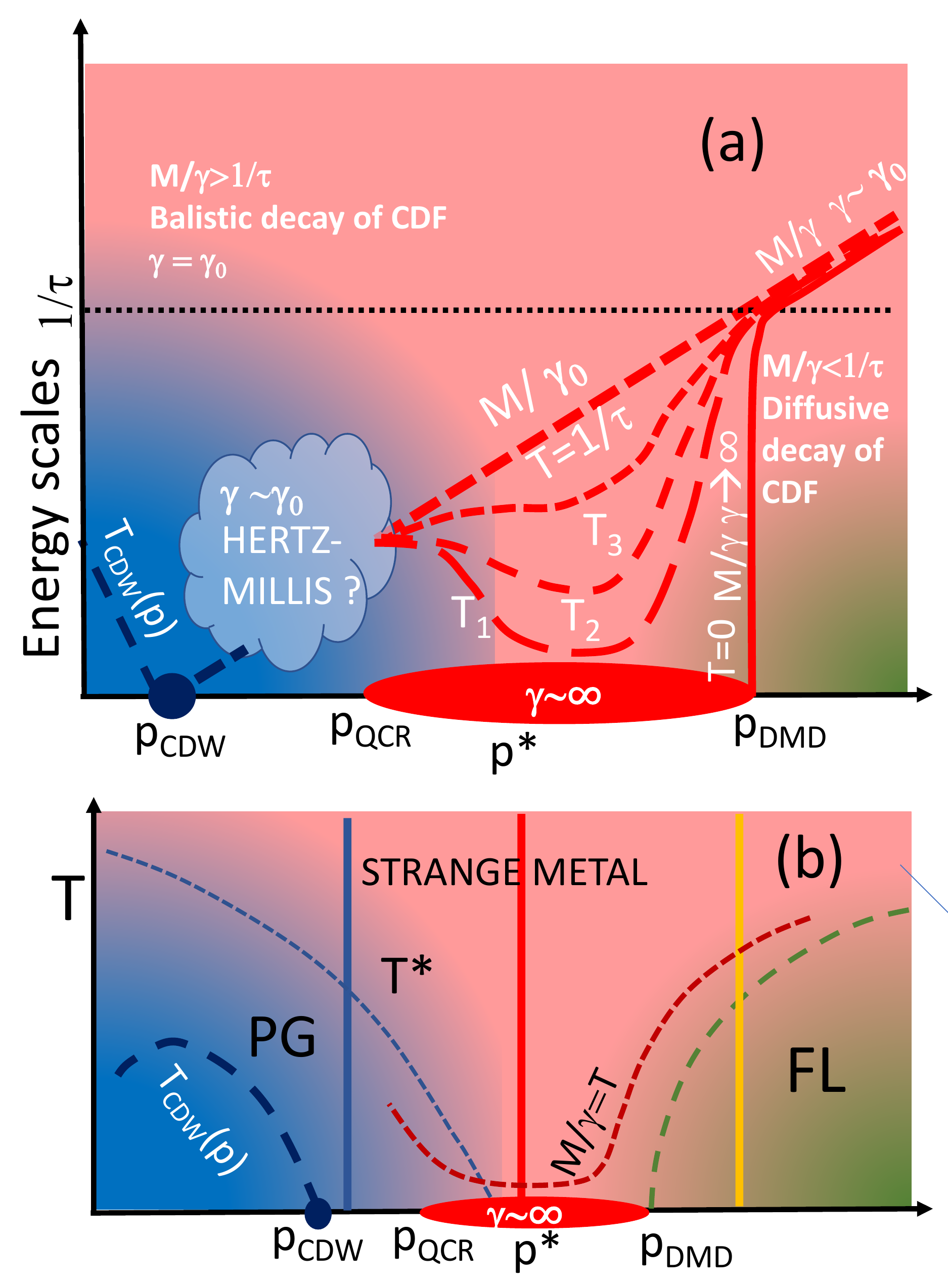}
\includegraphics[angle=0,scale=0.2]{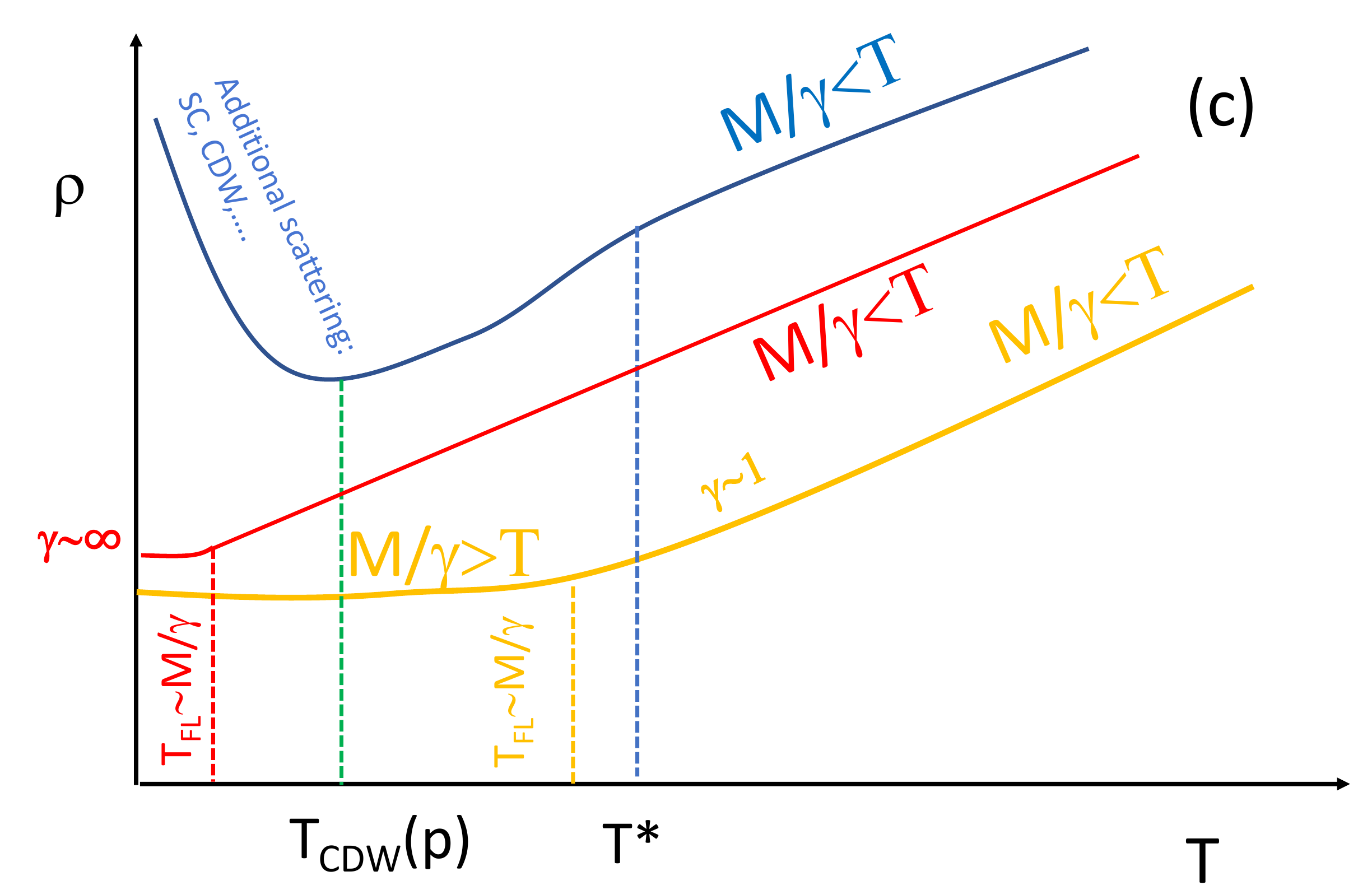}
\caption{(a) Schematic representation of the energy scales and their evolution with doping and temperature.  The green shaded area corresponds 
to the Fermi liquid region of the phase diagram (see panel (b)), the blue region is where the pseudogap is present, while in the reddish region 
the strange metal occurs. The dark blue dashed lines separate the standard renormalized classical, quantum critical and quantum 
disordered regions around the true CDW-QCP. The red dashed line mark the doping and temperature evolution of the characteristic energy scale
of CDF, $M/\gamma$, which decrease substantially when $M/\gamma$ becomes smaller than $1/\tau$ (dotted black line). 
In the present scenario the unknown connection between the large-$\gamma$ region and the standard Hertz-Millis QCP is represented by
a question mark in a cloud.
 (b) Schematic phase diagram of cuprates as a function of doping $p$ reporting the pseudogap crossover temperature 
$T^*$ (blue short-dashed line), the hidden transition line $T_{\mathrm{CDW}}(p)$ for charge density wave formation, ending at $T=0$ into a CDW-QCP at 
$p_{\mathrm{CDW}}$ is represented by a dark-blue long-dashed line.  Below the green dashed line the Fermi-liquid regime takes place. The red dashed line
marks the crossover $M/\gamma =T$ between the semiclassical region of CDF (above it) and the quantum regime (below it).
The yellow, red and blue vertical lines correspond to the resistivity vs temperature curves reported in (c).
(c) Schematic behavior of the resistivity as a function of $T$ at the doping values corresponding to the regions of the yellow, red 
and blue  vertical lines of panel (b).}
\label{fig-summary}
\end{figure}

{\it {--- The cuprates: resistivity ---}}
Regarding the transport properties, we showed in Ref.\,\onlinecite{seibold-2020} that CDFs account for the linear-in-$T$ 
resistivity in optimally and slightly overdoped YBCO and NBCO samples down to the superconducting critical temperature or slightly above it.  The 
question remained about the linear-in-$T$ resistivity observed for $p \approx p^*$ in strong magnetic fields suppressing superconductivity, 
which extends down to few K.  Our scenario is summarized in Fig.\,\ref{fig-summary}.
In Fig.\,\ref{fig-summary}(a) we schematically report with red dashed lines the behavior of the CDF characteristic energy at 
different temperatures $T_1<T_2<T_3<1/\tau$, both in the ballistic and in the diffusive regimes of their decay. This latter regime occurs when 
$M/\gamma_0<1/\tau$, which defines the doping $p_{DMD}$. At $T=0$, within our model, the characteristic energy drops to zero 
below $p_{\mathrm{DMD}}$, due to the logarithmic divergent $\gamma(T)$. For $p<p_{\mathrm{QCR}}$, the value of $\gamma$ is also
influenced by the pseudogap, possibly leading to $\gamma<1$ due to reduced phase space for damping processes. In any case, how 
$M/\gamma$ connects to the standard Hertz-Millis QCP is an open issue and therefore corresponds to {\it terra incognita} in 
Fig.\,\ref{fig-summary}(a). Since typical values for the scattering rates in cuprates are $1/\tau \sim 30-50$\,meV, one can notice that
the customarily reported phase diagrams of cuprates are usually in the regime where  $T<1/\tau$. Under this condition
we report in Fig.\,\ref{fig-summary}(b) a sketch of a cuprate phase diagram, where the red dashed line indicates the crossover 
temperature from the semiclassical to the quantum regime of CDFs, determined from $A\log[1/(\tau T)]=M/T$. It shows a 
significant drop below doping $p_{\mathrm{DMD}}$ [again due to the logarithmic $\gamma(T)$ behavior] and for the parameters given in 
Fig.\,\ref{fig-summary}, approaches a small but finite value $M/\gamma \lesssim 1$\,K in the doping range between pseudogap and FL region. 
In the resistivity and doping $p \gtrsim p_{\mathrm{QCR}}$ this reflects in the crossover from linear to quadratic behavior as shown in
Fig.\,\ref{fig-summary}(c), see also Refs.\,\onlinecite{seibold-2020,caprara-2022,mirarchi}. In the pseudogap region additional 
scattering mechanisms influence on $\rho(T)$, inducing a decrease from linearity below $T^*$ and an eventual increase at lower temperatures.

{\it {--- Discussion ---}}
In this work, we investigated the dynamics of OPF in the surrounding of a QCP. Although the quantum OPFs are intrinsically dynamical even 
at finite temperature, our results show that a divergent dissipation destroys this quantum character leading to fluctuations that are 
semiclassical down to $T=0$. This effect is similar to that found in Ref.\,\onlinecite{MMS-2002}, where dissipation quenches the 
instantons describing the quantum tunneling between local free energy minima of a disordered system. This classical statistics is then directly 
reflected in the linear-in-$T$ resistivity \cite{seibold-2020,caprara-2022,mirarchi}.

A few remarks are now in order. First of all, the ingredients of quenched impurities and of 2D short-ranged OPFs are 
so generic that a similar mechanism can easily be at work in other (maybe all) systems where the strange metal behavior is observed 
in the form of a linear-in-$T$ resistivity and a logarithmic $C_V/T$. The heavy fermion systems CeCu$_{6-x}$X$_x$ (X=Au, Ag) are 
just possible examples out of many others \cite{stewart,stockert-1998,coleman-2000}. The proximity to a QCP (charge density waves for 
cuprates, antiferromagnetic for heavy fermions) easily accounts for the observation of scaling properties. Our guess, within our scenario, 
is that scaling, being based on a large, diverging $\xi$, is truly observed at the QCP \cite{coleman-2000}, while the true strange 
metal behavior (with linear-in-$T$ resistivity) should occur away from it, where $M/\gamma$ is small or vanishing but $\xi$ 
is finite and still rather short. This mismatch between the precise tuning of $x=x_{\mathrm c}$ to observe scaling properties and bona-fide 
criticality at the QCP and the (possibly extended) range of strange metal behavior at $x_{\mathrm{QCR}}<x<x_{\mathrm{DMD}}$ 
(with $x_{\mathrm c}<x_{\mathrm{QCR}}$) is a definite testable prediction of our scenario, which calls for a more precise experimental 
scrutiny (actually this is the case in the YBa$_2$Cu$_3$O$_y$ cuprates, where $p_c=p_{\mathrm{CDW}}=0.16$, while strange metal properties are observed around 
$p^*\approx 0.19$ \cite{badoux-2016}).  Concerning cuprates, another intriguing, so far unsolved, issue concerns the effect of pseudogap 
and superconductivity on the dissipation parameter $\gamma$. Since this latter is naturally related to the density of states of particles 
near the Fermi surface (both in the ballistic and in the diffusive regimes), we argue that pseudogap and superconductivity should induce 
a decrease of $\gamma$ thereby opposing the strange metal behavior, as indeed observed below $T^*$.

Another prediction of our scenario is related to the $M/\gamma=1/\tau$ condition setting the doping regime where diffusion leads to 
an increasing $\gamma$. We suggest that an increase of disorder (e.g., by ion irradiation) might increase the elastic scattering rate 
$1/\tau$ extending the range in $T$ and $p$ where strange metal properties are observed.

A final remark is that the divergence of $\gamma$ at $T=0$ marks a complete slowing down of the OPFs, which acquires a vanishing 
characteristic energy. Considering an ensemble of OPFs which freeze when $\gamma \to \infty$, one might speculate that some kind of 
glassy state of frozen short-ranged OPFs might occur at $T=0$ over an extended range of $x$ slightly above a 2D QCP. This 
is another intriguing testable consequence of our scenario. 

{\it --- Acknowledgments ---}
We thank Riccardo Arpaia, Lucio Braicovich, Claudio Castellani, and Giacomo Ghiringhelli for  stimulating discussions.
We acknowledge financial support from the University of Rome Sapienza, through the projects, 
Ateneo 2019 (Grant No. RM11916B56802AFE), Ateneo 2020 (Grant No. RM120172A8CC7CC7), Ateneo 2021 (Grant No. RM12117A4A7FD11B), from the 
Italian Ministero dell'Universit\`a e della Ricerca, through the Project No. PRIN 2017Z8TS5B. G.S. acknowledges financial support
from the Deutsche Forschungsgemeinschaft under SE806/20-1.

\end{document}